\documentclass[11pt]{article}
\pdfoutput=1
\usepackage{jcappub,cancel}
\usepackage{paralist}
\usepackage{booktabs}
\usepackage[a4paper]{geometry}
\graphicspath{{images/}}
\usepackage{subfigure}
\usepackage{hyperref}

\title{An observable electron-positron anisotropy cannot be generated by
dark matter}

\author[a,b]{Stefano Profumo}
\affiliation[a]{Department of Physics, University of California, Santa Cruz\\1156 High St, Santa Cruz, CA 95064}
\affiliation[b]{Santa Cruz Institute for Particle Physics,\\ 1156 High St, Santa Cruz, CA 95064}
\emailAdd{profumo@ucsc.edu}

\begin{document}

\abstract{ I demonstrate that if an anisotropy in the arrival direction of high-energy cosmic-ray electrons and positrons is observed then, barring local anisotropic diffusion,  dark matter annihilation is ruled out as an explanation to the positron excess. For an observable anisotropy to originate from dark matter annihilation, the high-energy electrons and positrons must be produced in a nearby clump. I consider the annihilation pathway producing the smallest flux of gamma rays versus electrons and positrons, and the combination of clump distance and luminosity that minimizes the gamma-ray flux. I show that if an anisotropy from such a clump were detected, and if such anisotropy did not generate from anisotropic diffusion effects, then the clump would be clearly detectable as an anomalous, bright gamma-ray source with the Fermi Large Area Telescope. I also point out that the non-detection of an anisotropy is perfectly compatible with an astrophysical origin for the excess positrons that has nothing to do with dark matter.}

\maketitle

\section{Introduction}
A rising positron fraction at energies of 10 GeV and above, tentatively observed more than two decades ago \cite{earlypepm}, has now been well established by measurements with PAMELA \cite{pamela}, the Fermi Large Area Telescope (LAT) \cite{latepem} and, recently, AMS-02 \cite{ams}.  The fraction of cosmic-ray positrons to electrons-plus-positrons declines up to about 7 GeV, and is observed, with high statistics, to increase to the level of 0.1 at 100 GeV and about 0.15 all the way up to 350 GeV \cite{ams}.

A declining positron fraction in the energy range above $\sim1$ GeV is a generic and well-established prediction of diffusive models for Galactic cosmic ray propagation in the absence of nearby primary sources of positrons \cite{Moskalenko:1997gh}. Measurements of the total electron-positron flux and of the positron fraction are spectrally compatible with the existence of one or more additional primary sources of electrons and positrons \cite{latelectron, Grasso:2009ma}. The nature of such additional primary source(s) remains, however, elusive.

Numerous studies have attributed the excess primary positrons to the pair-annihilation of Galactic dark matter (DM), a scenario strongly constrained by prompt and secondary radiation produced in the annihilation events, but at present still consistent with current observations (see e.g.  Ref.~\cite{Cholis:2013psa}). Observations are also fully consistent with primary cosmic-ray electrons and positrons produced in the magnetosphere of nearby mature pulsars (see e.g. Ref.~\cite{Linden:2013mqa} for a recent discussion of a single pulsar origin, and \cite{DiMauro:2014iia} for multiple pulsars). In addition, the possibility of {\em in situ} acceleration has also been entertained \cite{Mertsch:2014poa}, a scenario soon to be tested with high-energy observations of the boron-to-carbon ratio \cite{Cholis:2013lwa}. Finally, drastic departures from a diffusive propagation picture might also reconcile observations with a purely secondary positron origin \cite{Cowsik:2013woa}.

A possible test of the origin of the excess high-energy positrons is the search for an anisotropy in the arrival direction of cosmic ray electrons and positrons. Albeit such charged particles' trajectories are bent as they meander through the Galactic magnetic fields, if indeed they originate from a nearby accelerator a residual, and potentially detectable, anisotropy might be observed (see e.g. Ref.~\cite{Hooper:2008kg, Profumo:2008ms, Borriello:2010qh}). Given the energy loss timescale for electrons and positrons with energies in the hundreds of GeV, such particles need to be produced within a few kpc of the Sun's position in the Galaxy. As a result, DM annihilation in the center of the Galaxy does not significantly contribute to local high-energy cosmic-ray electrons and positrons. It is well-established that the corresponding level of anisotropy is negligible, and well below detectable levels, independent of the choice for the (smooth) DM density profile \cite{Cernuda:2009kk, Borriello:2010qh}. The only possible scenario for DM annihilation to produce a detectable anisotropy is in the presence of a large, nearby clump, as first proposed in Ref.~\cite{Hooper:2008kv} (see also Ref.\cite{brunetal, Cernuda:2009kk, Borriello:2010qh}). It is crucial, however, to entertain the possibility that an anisotropy be generated by anisotropic diffusion of cosmic rays, see e.g. the discussion in Ref.~\cite{Borriello:2010qh} and the recent results of Ref.~\cite{Mertsch:2014cua} and references therein.

It has been shown that the likelihood for the existence of a luminous enough DM clump to produce the excess positron fraction is generically quite low \cite{brunetal}, depending on the DM particle mass, annihilation final state and pair-annihilation rate. Ref.~\cite{brunetal} also pointed out that in almost any instance where the excess positrons are produced by DM annihilation in a nearby clump, such clump would likely be  detectable in gamma rays.

Here, I generalize the findings of Ref.~\cite{brunetal} and demonstrate, in an entirely analytic way, that a DM clump responsible for the excess positrons, and producing a detectable electron-positron cosmic-ray anisotropy, necessarily produces a gamma ray flux much brighter than the LAT 5$\sigma$ point source sensitivity. Given the absence, in the LAT data, of such a bright, unidentified gamma-ray source, presumably with a characteristic spectral shape reminiscent of DM annihilation, this result shows that the detection of an anisotropy would eliminate DM annihilation as the primary explanation for the anomalous rise in the positron fraction. I also show that the DM clump that would best ``hide'' in gamma rays is extraordinarily unlikely to exist given the results of N-body simulations of Galactic DM halos \cite{vl2}.

The sketch of the ensuing proof is as follows: 

(i) I consider the DM annihilation final state that produces the smallest possible amount of gamma rays per cosmic-ray electrons and positrons (a monochromatic electron-positron pair); 

(ii) I analytically solve the diffusion and energy-loss propagation equation; 

(iii) I impose that the clump produce the observed positron fraction, and 

(iv) I calculate the minimal, guaranteed (internal bremsshtrahlung) gamma-ray emission from the clump; 

(v) I show that such minimal emission is inversely proportional to the cosmic-ray anisotropy, and that for detectable anisotropies it is more than an order of magnitude larger than the LAT point-source sensitivity; finally, I show that 

(vi) a clump at the optimal distance and luminosity to suppress gamma-ray emission has a likelihood of existence of roughly one part in $10^4$, according to N-body simulations.

\section{Dark matter clumps, cosmic-ray anisotropy and gamma-ray fluxes}

I define a DM clump luminosity, with units of inverse time, as 
\begin{equation}
{\cal L}\equiv\frac{\langle\sigma v\rangle}{2m_\chi^2}L_{\rm clump},
\end{equation}
with, as customary, $\langle\sigma v\rangle$ the thermally-averaged, zero-temperature pair-annihilation cross section times relative velocity, $m_\chi$ the DM particle mass, and
$$
L_{\rm clump}\equiv\int_{\rm clump}\rho_{\rm DM}^2\ {\rm d}^3x,
$$
with the integral running over the volume of the DM clump, and $\rho_{\rm DM}$ the clump DM density \cite{brunetal}.

In the diffusive propagation picture, the average dipolar anisotropy in the direction of a source (in this case the DM clump) versus the opposite direction, at cosmic-ray electron-positron ($e^\pm$)  energy $E$ is given by \cite{1972ChJPh..10...16M, Ackermann:2010ip}
\begin{equation}\label{aniso}
\Delta(E)\equiv\frac{I_{\rm max}-I_{\rm min}}{I_{\rm max}+I_{\rm min}}\simeq\frac{\phi_{e,{\rm clump}}(E)}{\phi_{e,{\rm TOT}}(E)}\frac{3D(E)}{c}\frac{2d}{\lambda(E)},
\end{equation}
with $\phi_{e,{\rm clump}}(E)$ the flux of $e^\pm$ from the clump, $\phi_{e,{\rm TOT}}(E)$ the total $e^\pm$ flux, $$D(E)=D_0\left(\frac{E}{E_0}\right)^\delta$$ the diffusion coefficient, $d$ the distance to the clump, with the exponent $\delta\simeq0.7$ capturing the energy dependence of the diffusion coefficient, and $\lambda(E)$ the diffusive ``area'' associated with $e^\pm$ propagation from the clump, to be specified below.

Note that the ratio
\begin{equation}
\frac{\phi_{e,{\rm clump}}(E)}{\phi_{e,{\rm TOT}}(E)}\simeq 2P_{e^\pm}(E),
\end{equation}
with $P_{e^\pm}(E)$ the positron fraction at energy $E$, where I assume that the clump sources most of the excess positrons at energy $E$, a condition that maximizes the observable cosmic-ray anisotropy.

In  what follows, I calculate the anisotropy at the largest possible energy where significant statistics and a robust constraint can be obtained with current observations. As shown in previous work \cite{Profumo:2008ms, Cernuda:2009kk, Ackermann:2010ip}, the largest anisotropy is in fact associated with the largest possible energies (below the DM particle mass). While the Fermi-LAT sensitivity to $e^\pm$ anisotropies  worsens at increasing energy, the best constraints on cosmic-ray anisotropy with the LAT have also been placed at the largest accessible energies \cite{Ackermann:2010ip}. The AMS-02 constraints on anisotropy are flat in energy and constrain $\Delta<0.036$, thus the larger the energy the stronger the constraints \cite{ams}. For definiteness, here I pick $E=300$ GeV, close to the largest energy probed by AMS-02 \cite{ams}. I will argue below that this is actually a conservative choice (i.e., I obtain even larger gamma-ray fluxes for $E<300$ GeV).

In order to obtain the smallest possible gamma-ray flux from the DM clump producing the excess positrons, one needs to consider the annihilation final state producing the smallest possible gamma-ray per electron-positron yield. This corresponds to DM promptly annihilating to a (monochromatic) $e^\pm$ pair. Hadronic final states, or gauge boson pairs decaying hadronically, produce copious gamma rays from pion decay, and heavier leptons also produce significantly more gamma radiation than monochromatic $e^\pm$. Annihilation to non-standard model particles subsequently decaying into lower-energy $e^\pm$ would yield even more gamma rays from the same internal bremsstrahlung process than the monochromatic case. To be additionally conservative, I also neglect secondary inverse Compton or secondary bremsstrahlung emission, and exclusively consider internal bremsstrahlung as the only source of gamma radiation from the DM clump.

I now estimate the flux of $e^\pm$ and of gamma rays from a given clump with luminosity $\cal L$ at a distance $d$. Once integrated over time for the case of a stationary source such as a DM clump, the Green's function of the diffusion equation for the $e^\pm$ distribution function $f(\vec r, E, t)$
\begin{equation}\label{diffeq}
\frac{\partial f}{\partial t}-D(E)\Delta f+\frac{\partial}{\partial E}\left(b(E)f\right)=Q(\vec r, E, t)
\end{equation}
directly gives the solution for the present case. The ``source term'' for the Green's function is of course a delta function in space, time and energy, $$Q(E)\sim\delta(\vec r-\vec r_0)\delta(t-t_0)\delta(E-m_\chi),\ \ {\rm with}\ \ |\vec r-\vec r_0|=d,$$  where $b(E)=b_0(E/E_0)^2$, with $b_0\sim10^{-16}$ GeV/s and $E_0\simeq 1$ GeV. The Green's function for the differential equation (\ref{diffeq}) is known \cite{1959SvA.....3...22S}. The resulting particle flux (before solar modulation, which is entirely irrelevant for the energies under consideration here, and in the case of a stationary process) reads (see Eq.~(11) of Ref.~\cite{1959SvA.....3...22S}):
\begin{equation}\label{clumpe}
\phi_{e,{\rm clump}}(E)=\left[\frac{1}{b(E)}\cdot\frac{\exp\left(-\frac{d^2}{4\lambda(E)}\right)}{(4\pi\lambda(E))^{3/2}}\right]\cdot\frac{c}{4\pi}\cdot{\cal L},
\end{equation}
with 
\begin{equation}\label{lambda}
\lambda(E)=\frac{D_0E_0}{b_0(1-\delta)}\left[\left(\frac{E_0}{E}\right)^{1-\delta}-\left(\frac{E_0}{m_\chi}\right)^{1-\delta}\right].
\end{equation}
Notice that 
\begin{equation}\label{upperlambda}
\lambda<\frac{D_0E_0}{b_0(1-\delta)}\simeq 3\times10^{44}\ {\rm cm}^2.
\end{equation} 
In what follows, one must ensure that this condition be self-consistently fulfilled.

The  differential gamma-ray flux associated with the DM clump where DM particles annihilate into $e^\pm$ pairs, stemming from internal bremsstrahlung only, is (see e.g. \cite{Beacom:2004pe})
\begin{equation}\label{clumpgr}
\phi_{\gamma}=\frac{\cal L}{4\pi d^2}\frac{4\alpha}{\pi}\frac{\ln\left(2\frac{m_\chi}{m_e}\right)}{E_\gamma},
\end{equation}
with the integrated flux above some threshold energy $E_{\gamma,0}$
\begin{equation}\label{clumpgrint}
\phi_{\gamma,{\rm TOT}}=\frac{\cal L}{4\pi d^2}\frac{4\alpha}{\pi}\ln\left(2\frac{m_\chi}{m_e}\right)\ln\left(\frac{m_\chi}{E_{\gamma, 0}}\right).
\end{equation}
Let us cast the anisotropy of Eq.~(\ref{aniso}), calculated at $E=300$ GeV, as
\begin{equation}\label{deltaaniso}
\Delta=\frac{d\cdot L}{\lambda},
\end{equation}
with the quantity $L$, with dimensions of length, defined as
$$
{L}\simeq {3\times10^{19}}{{\rm cm}}\left(\frac{2P_{e^\pm}(E=300\ {\rm GeV})}{0.3}\right)\left(\frac{D_0}{10^{28}\ {\rm cm}^2{\rm s}^{-1}}\right),$$
and where I used $\delta=0.7$.
Inferring the clump cosmic-ray $e^\pm$ flux from the measured $e^\pm$ flux and the measured positron fraction, I get a luminosity
\begin{equation}
{\cal L}\simeq\frac{9\times10^{-30}}{{\rm cm}^3\ {\rm s}}\frac{\lambda^{3/2}}{\exp(-d^2/(4\lambda))},
\end{equation}
where, as a reminder, $\lambda$ has units of length squared.
The resulting differential gamma-ray flux reads, for  $m_\chi\sim{\cal O}(1\ {\rm TeV})$ (note that the dependence on mass is only logarithmic), and for $E_\gamma=10$ GeV
\begin{equation}
{\phi_\gamma}{}\simeq {10^{-2}}\frac{\cal L}{d^2}\simeq\frac{10^{-31}}{{\rm cm}^3\ {\rm s}}\frac{\lambda^{3/2}}{d^2\exp(-d^2/(4\lambda))}.
\end{equation}
Using Eq.~(\ref{deltaaniso}) above,  I get
\begin{equation}\label{eq3}
{\phi_\gamma}\simeq\frac{10^8}{{\rm cm}\ {\rm s}}\frac{1}{\Delta^2\sqrt{\lambda}\exp(-\Delta^2\lambda/(4L^2))}.
\end{equation}
The gamma-ray flux is minimized when the function of $\lambda$ in the denominator is maximized, which happens for $$\lambda_{\rm max}=4L^2/\Delta^2.$$ As a consistency check, note that the corresponding value of $\lambda$ is compatible with the upper limit I obtained before in Eq.~(\ref{upperlambda}): $$\lambda_{\rm max}\simeq4\times 10^{43}{\rm cm}^2\left(\frac{L}{3\times 10^{19}{\rm cm}}\right)^2\left(\frac{10^{-2}}{\Delta}\right)^2<\frac{D_0E_0}{b_0(1-\delta)}.$$
Substituting for $\lambda_{\rm max}$ in Eq.~(\ref{eq3}) I find, for the differential gamma-ray flux at $E_\gamma=10$ GeV,
\begin{equation}\label{finaleq}
\phi_\gamma>\left(\frac{10^{-2}}{\Delta}\right) \frac{5\times 10^{-10}}{{\rm GeV}\ {\rm cm}^2\ {\rm s}}\gg \phi_\gamma^{{\rm Fermi},5\sigma}\simeq\frac{{\rm few}\times 10^{-11}}{{\rm GeV}\ {\rm cm}^2\ {\rm s}},
\end{equation}
implying a gamma-ray flux well above the 5$\sigma$ Fermi LAT point source sensitivity \cite{fermisens} at 10 GeV, which ranges from $3\times10^{-11}/({\rm GeV}\ {\rm cm}^2\ {\rm s})$ for a high-latitude source, to $\sim10^{-10}/({\rm GeV}\ {\rm cm}^2\ {\rm s})$ for a source on the Galactic plane. Clearly, for a large enough anisotropy $\Delta \sim10^{-2}$, i.e. at a detectable level with AMS or with Fermi, the gamma-ray flux is bright enough to be solidly detectable by the Fermi LAT. Notice that the current AMS limits already imply $\phi_\gamma>1.4\times10^{-10}/({\rm GeV}\ {\rm cm}^2\ {\rm s})$, at the Fermi LAT sensitivity even for a source in the Galactic plane.

A similar calculation for the integrated flux above $E_{\gamma,0}=0.1$ GeV, and again for $m_\chi=1000$ GeV, yields identical conclusions\footnote{Note that the integral sensitivity for Fermi is usually calculated for a $1/E^2$ spectrum, while here we have a $1/E$ spectrum; this does not affect quantitatively the results presented here.}:
\begin{equation}\label{intflux}
\phi_{\gamma,{\rm TOT}}>\left(\frac{10^{-2}}{\Delta}\right) \frac{4\times 10^{-8}}{{\rm cm}^2\ {\rm s}}\gg \phi_{\gamma,{\rm TOT}}^{{\rm Fermi},5\sigma}\simeq\frac{{\rm few}\times 10^{-9}}{{\rm cm}^2\ {\rm s}}.
\end{equation}
Notice that what I obtained above would only be strengthened by choosing a lower value for the $e^\pm$ energy $E$ (for example, the gamma-ray fluxes in Eq.~(\ref{finaleq}) and (\ref{intflux}) would be a factor of about 2 larger for $E=30$ GeV, and Fermi-LAT data constrain the anisotropy to be even smaller than $10^{-2}$ at energies below 100 GeV \cite{Ackermann:2010ip})

One can easily infer the distance and luminosity of the DM clump invoked above which minimizes the gamma-ray flux while producing the excess positrons and a detectable level of electron-positron anisotropy. I find
\begin{equation}
d_{\rm clump}\simeq4\ {\rm kpc}\left(\frac{\Delta}{10^{-2}}\right),
\end{equation}
and
\begin{equation}
L_{\rm clump}\simeq7\times 10^{10} \frac{M_\odot^2}{{\rm pc}^3}\left(\frac{m_\chi}{1\ \rm {TeV}}\right)^2\left(\frac{3\times 10^{-26}{\rm cm}^3/{\rm s}}{\langle\sigma v\rangle}\right).
\end{equation}
According to the numerical results of the Via Lactea-II simulation \cite{vl2} for the distribution of DM clumps in a Milky Way-type galactic halo, the likelihood of having a clump at that distance and with that luminosity is on the order of 0.01\% (see fig.~2 of \cite{brunetal}). Such likelihood can be increased for larger pair-annihilation rates, but it  would still be very small for any phenomenologically acceptable value of the pair-annihilation cross section. A similar conclusion emerges from analytical considerations along the lines of Ref.~\cite{Kamionkowski:2010mi}

\section{Discussion and Conclusions}
In the usual diffusion scheme for the propagation of Galactic electron-positron cosmic rays, I related the dipolar anisotropy $\Delta$ for the cosmic-ray arrival direction from a putative DM clump sourcing the excess positrons, to a minimal, guaranteed associated DM clump gamma-ray luminosity. I chose the most conservative possible setup, meaning the DM annihilation final state producing the smallest amount of gamma rays, and I neglected secondary radiation. Even with these conservative assumptions, I demonstrated with Eq.~(\ref{finaleq}) that for any anisotropy $\Delta$ large enough to be detectable, the clump should be very bright in gamma rays, and well above the Fermi LAT point source sensitivity.

Since no bright, unassociated gamma-ray source has been found with a spectrum that could originate from DM \cite{nofermiunassociated}, the present result implies that the detection of an anisotropy in the cosmic-ray electron-positron arrival direction would rule out DM as the source of the excess positrons. Note that a possible caveat to this conclusion is a local magnetic field structure that could funnel otherwise close-to-isotropically distributed cosmic-ray electrons and positrons to produce an ``artificial'' and otherwise undetectable anisotropy.

One might wonder which implications the findings presented here have for astrophysical sources of the excess cosmic-ray positrons, such as pulsars or supernova remnants. On the one hand, while the results presented here indicate that the detection of an anisotropy would eliminate DM as the explanation to the excess cosmic-ray positrons, such anisotropy is not guaranteed from astrophysical sources. Simple counter-examples include the presence of more than one astrophysical source contributing to the excess positrons, or local magnetic field structures reshuffling the cosmic-ray trajectories so as to erase any original directionality (in practice invalidating Eq.~(\ref{aniso}).

It is important to emphasize that the results presented here depend on the assumption of isotropic and spatial uniformity of cosmic ray diffusion. Local magnetic field turbulence can disrupt significantly such assumption. Hints of such a possibility have appeared in hadronic cosmic rays at large energies (larger than 10 TeV), see e.g. Ref.~\cite{Giacinti:2011mz}. It is, however, hard to infer from such observations the level of possible effects on anisotropy at lower energy, and for leptonic cosmic rays. Yet, this is a key caveat that the Reader should bear in mind.

A second caveat is the possibility that a dark matter clump resided so close to us that the assumption of diffusive behavior of the electrons and positrons produced by dark matter annihilation, Eq.~(\ref{aniso}), would not be valid. A similar possibility is that the Sun resides inside a local dark matter over-density large enough to impact the anisotropy levels predicted here. Again, the diffusive behavior assumed above would not be established and the conclusions presented here would not directly apply. 

One might also wonder if these results can be applied to astrophysical sources: should an anisotropy be observed, do we expect the associated astrophysical source to be bright in gamma rays? The answer is likely yes, since for any reasonable astrophysical source the associated gamma-ray brightness will be larger than what conservatively considered here. Of course, most of the relevant local astrophysical sources are in fact well-established gamma-ray sources. One subtlety is that the injection time for DM is constant in time, while e.g. for a pulsar the bulk of the electrons and positrons is injected at one given point in time. What we derived here might be useful (when suitably modified for the relevant electron-positron spectrum expected from a given astrophysical source, and for the injection time, e.g. associated with a pulsar's age) to predict the ballpark of the expected gamma-ray emission, should the detection of a significant anisotropy occur.

\section*{Acknowledgments}
%
\noindent This work is partly supported by the US Department of Energy, Contract DE-FG02-04ER41268. 
%

\end{document}